\documentclass[twocolappendix,numberedappendix]{emulateapj}

\newcommand{\LCDM}{$\Lambda$CDM }

\newcommand{\kms}{km $s^{-1}$}
\newcommand{\hinv}{h^{-1}}
\newcommand{\mpc}{\rm{Mpc}}
\newcommand{\hmpc}{\hinv\mpc}
\newcommand{\kpc}{\rm{kpc}}
\newcommand{\hkpc}{\hinv\kpc}

\newcommand{\hmsol}{\hbox{$\hinv {\rm M}_\odot$}}

\shorttitle{Subhalo accretion through filaments}
\shortauthors{Gonz\'alez \& Padilla}

\begin{document}

\title{Subhalo accretion through filaments}
\author{Roberto E. Gonz\'alez$^{1,2}$ and Nelson D. Padilla$^{1,2}$}
\affil{$^1$ Instituto de Astrof\'{i}sica, Pontificia Universidad Cat\'olica,
  Av. Vicu\~na Mackenna 4860, Santiago, Chile\\
  $^2$ Centro de Astro-Ingenier\'{i}a, Pontificia Universidad Cat\'olica,
  Av. Vicu\~na Mackenna 4860, Santiago, Chile\\
}
\email{regonzar@astro.puc.cl}

\begin{abstract}

We track subhalo orbits of galaxy and group sized halos in cosmological simulations.
We identify filamentary structures around halos and we use these to define 
a sample of subhalos accreted from filaments as well as a control sample of subhalos accreted from other directions. 
We use these samples to study differences in satellite orbits produced by filamentary accretion.
Our results depend on host halo mass.  We find that for low masses, subhalos accreted from filaments show $\sim10\%$ shorter lifetimes compared to the control sample, { they show a tendency towards more radial orbits}, reach halo central regions earlier, and are more likely to merge with the host.
For higher mass halos
this lifetime difference dissipates and even reverses for cluster sized halos.
This behavior appears to be connected to the fact that more massive hosts are connected to stronger filaments with higher velocity coherence and density, with { slightly} more radial subhalo orbits. Because subhalos tend to follow the coherent flow of the filament, it is possible that such thick filaments are enough to shield the subhalo from the effect of dynamical friction at least during their first infall.
We also identify subhalo pairs/clumps which merge with one another after accretion. 
They survive as a clump for only a very short time, which is even shorter for higher subhalo masses,
suggesting that the Magellanic Clouds and other Local group satellite associations, may have entered the MW virial radius very recently and probably are in their first infall.

\end{abstract}


\section{Introduction}

Naively, one would expect that in the \LCDM cosmology, the properties of the satellite population depend only on host halo mass.
In particular, there are several studies which indicate that the halo occupation distribution at fixed halo mass is statistically independent of the large-scale environment \citep{1991ApJ...379..440B, 1999MNRAS.302..111L, 2004ApJ...609...35K}.

However, { already the early work by \citet{1997MNRAS.290..411T} showed that the masses and orbits of satellites accreted by central clusters was dependent on the cluster mass, and highly anisotropic. This evolved into more recent results}  obtained using both numerical simulations and observations which indicate that the satellite population may be affected by the large-scale environment, particularly their dynamics and locations.
\citet{2013ApJ...770...96G} found an enhancement in subhalo velocity for galactic hosts with a M31-like companion, when compared to an isolated control sample of the same mass. In SDSS, galaxies in filaments have more satellites \citep{2015ApJ...800..112G}, and satellites tend to align with the filaments\citep{2015MNRAS.450.2727T}.

Another important issue related with subhalo orbits is the incidence of subhalo { structures such as discs, and the later abundance of} pairs and clumps and their fate after accretion. 
{  \citet{2005MNRAS.363..146L} show that the incidence of disks or planes of satellites is to be expected to some degree in a $\Lambda$CDM cosmology, in particular when considering the brightest subhalos.  Furthermore, the distribution of these bright satellites lies in a plane perpendicular to the main galaxy disc, in agreement with hints from the Milky Way \citep{2007MNRAS.374...16L}.  The probable reason for the formation of these discs of satellites is the accretion of subhaloes from filaments \citep{2011MNRAS.413.3013L}.  However this is still subject to debate as, e.g. \citep{2012MNRAS.424...80P}, it is pointed out that the planes of satellites in the Aquarius and Via Lactea simulations are not as flat as claimed by observational measurements for the Milky Way.

Regarding clumps , it is worth noting that the} Local Group contains several satellite associations in M31 and the MW; the Magellanic Clouds (MCs), for example, are an interacting 
satellite pair. 
In general these associations are rare in the \LCDM model, 
and little we know about their accretion history and evolution in a cosmological context. 
{ But this has also been associated to the existence of planes of satellites.  \citet{2008MNRAS.385.1365L} showed that the infall of satellites that were part of a clump before accretion could indeed produce a final distribution of satellites in the new host, distributed along a disc, with some remnant clumps.}
{ In addition, satellite-satellite mergers after infall are associated to satellites accreted in clumps with correlated infall histories. \citep{2014ApJ...794..115D,2015ApJ...807...49W}.  }

\citet{2013MNRAS.431L..73F} found that, in the Local group, $30\%$ of all MW and M31 satellites are found in likely physical pairs of comparable luminosity, and they are much closer together than expected by chance if the radial and angular distributions of satellites are uncorrelated. In contrast, for the same pair criteria, this fraction drops to less than $4\%$ in N-body/semi-analytic models that match the radial distribution and luminosity function of Local Group satellites.

Recent proper motion measurements of the Magellanic Clouds \citep{2006ApJ...638..772K, 2013ApJ...764..161K}, and their implied orbits, favor a first infall scenario \citep{2007ApJ...668..949B, 2010ApJ...721L..97B, 2013ApJ...764..161K}.
\citet{2011MNRAS.414.1560B} used numerical simulations to conclude that the LMC was accreted within the past four Gyr and is currently making its first pericentric passage about the MW; however, they consider the LMC and SMC as independent objects.

Massive satellites do not survive for long. Dynamical friction time-scales for $1:10$ objects at $z=1$ is $\sim 5$ $\rm{Gyr}$ \citep{2008MNRAS.383...93B}. \citet{2011ApJ...743...40B} used simulations to study the accretion history of the LMC and SMC as independent objects in the past $10$ $\rm{Gyr}$, and found a $72\%$ probability that the MCs were accreted in the last Gyr, $50\%$ of which were accreted as a pair.
In simulations, \citet{2013ApJ...770...96G} found that only $2\%$ of MW-sized halos have a subhalo pair($V_{circ} > 50$\kms), and only $1$ out of $30000$ have a MC-like pair.

The main goal of this paper is to understand the effect of filamentary accretion on the subhalo population and, as a secondary goal, to study the orbital evolution of subhalo associations (pairs/clumps) to try and use this to infer the accretion history of MCs and/or other satellite associations in the Local Group.

\section{Data}

\subsection{Simulations}

We use two sets of simulations.  One designed to study the orbits of accreted satellites in galaxy and small group sized haloes, and another for a cluster sized halos.

The simulation for smaller haloes, which will be referred to as FORS (Following ORbits of Satellites) from this point on,
was run using the {\small GADGET2} code \citep{2005MNRAS.364.1105S} with cosmological parameters consistent with those from the Planck mission \citep{2014A&A...571A..16P}, with matter, cosmological constant and baryon density parameters $\Omega_m=0.3175$, $\Omega_{\Lambda}=0.6825$, $\Omega_b=0.049$, and Hubble constant $H_0=100 h$km$/$s/Mpc, where $h=0.6711$.
The simulation contains $1024^3$ particles in a $40\hmpc$ side box. The particle mass is $5.25 \times 10^6$\hmsol.

A total of $301$ snapshots were stored, spaced by $\Delta a=0.005$ from scale factor $a=0.1$ to $a=0.6$, and by $\Delta a=0.002$ from $a=0.6$ to $a=1$, which results in a time spacing of $\sim 30$Myr at $z~0$.  This provides an appropriate time resolution to track accurate orbits for hundreds of MW-sized halos up to $M_{sub}/M_{host} \sim 10^{-4}$.

We use the Phoenix simulation suite \citep{2012MNRAS.425.2169G} for cluster sized halos.
In particular we choose the { Phoenix C and E simulations (Ph-C and Ph-E hereafter) which follow  clusters of final masses $\sim 5.5 \times 10^{14}$\hmsol $\,$ and $\sim 6 \times 10^{14}$\hmsol $\,$ respectively. For these simulations the particle mass is $4.43 \times 10^6$\hmsol. The number of snapshots in both cases is $72$ spaced by $\Delta a=0.017$. The mass resolution is similar among all simulations, but the orbits in the Phoenix simulations are more coarsely sampled. }

\subsection{Hosts}
Halos and merger trees in FORS are obtained using the Rockstar Halo Finder and Consistent Tree codes\citep{2013ApJ...762..109B}.
There are $\sim 1$ million halos with masses above $10^8$\hmsol; the mass of the largest halo is $3 \times 10^{14}$\hmsol. Around $20\%$ 
of all objects identified are subhalos.
The host halos for which we choose to track their subhalo orbits are intended to include masses encompassing hosts of the MW or M31  
up to small groups, i.e.,
\begin{itemize}
\item Mass at $z=1$ in the range $10^{11}$\hmsol$<M_{VIR}<10^{13}$\hmsol
\item mass at $z=0$ in the range $10^{12}$\hmsol$<M_{VIR}<10^{13}$\hmsol
\item Merger ratios less than $1:4$
\end{itemize}

\begin{figure}[!htb]
\begin{center}
\includegraphics[width=.95\linewidth,angle=0]{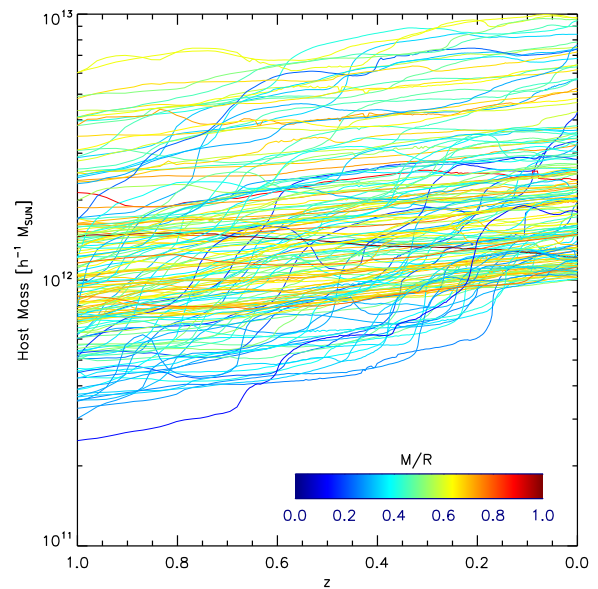}
\caption{
\label{fighost} Host halo mass evolution since $z=1$ for our selected sample in the FORS simulation.  The color scale indicates the mass growth ratio between $z=1$ and $z=0$. Most halos in the sample show a smooth mass evolution to avoid subhalo orbits affected by major mergers.}
\end{center}
\end{figure}

The resulting sample consists of $162$ host halos. Figure \ref{fighost} shows the mass evolution for the selected hosts since $z=1$; 
the color scale represents the mass growth ratio(M/R) since $z=1$ down to the present time.

Most host halos show a smooth mass evolution since $z=1$, ensuring we exclude systems where the subhalo population 
may be strongly altered by major mergers.
We also set this set of conditions in order to allow a sample of hosts comparable to those of the MW or M31.

To study the subhalo orbits of more massive halos, { we use the Ph-C and Ph-E simulations of clusters with masses  $\sim M_{200} \sim 6 \times 10^{14}$\hmsol. Notice that the Phoenix simulations adopt a $M_{200}$ definition for halo mass, instead of the virial mass used in FORS.}  
The former is generally smaller.
Subhalos are also found differently, with the SUBFIND code \citep{2001NewA....6...79S}; details about this cluster resimulation can be found in \citet{2012MNRAS.425.2169G}.
We do not expect large differences in the subhalo detection between Rockstar and SUBFIND \citep{2011MNRAS.415.2293K}.

\subsection{Subhalos}
We track the subhalo orbits of the selected hosts in both simulations.  We only consider subhalos which entered 
the virial radius after $z=1$ for the first time, i.e., our sample contains no subhalos that were inside the virial radius before that time. 
We do not adopt the subhalo tag provided by the halo finder algorithm. Instead, we define as subhalos all halos crossing the host virial radius.

\begin{figure}[!htb]
\begin{center}
\includegraphics[width=.95\linewidth,angle=0]{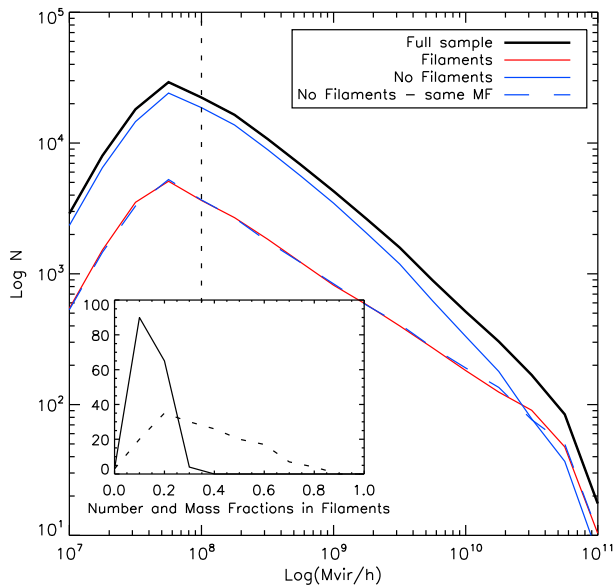}
\caption{
\label{fig1} Subhalo mass function at accretion time. Full sample (black), subhalos accreted through filaments (red), not from filaments (blue), and the No Filament sample, consisting of subhalos not coming from filaments but having the same mass function as from filaments (dashed blue). 
The bottom-left sub-panel shows the distribution of the number (solid) and mass (dashed) fraction of subhalos accreted from filaments in each host; on average we have $20\%$ subhalos accreted from filaments, but they account for $\sim40\%$ of the total accreted mass.}
\end{center}
\end{figure}

We find $125,335$ subhalos in the FORS simulation. Figure \ref{fig1} shows their mass function at accretion time.
For the remainder of this paper we select subhalos with masses above \hbox{$1.05 \times 10^8$\hmsol} or $20$ particles (vertical dashed line) and those which survive at least $4$ snapshots after accretion.  These cuts result in a final sample of $61896$ subhalos.

{
In the Ph-C and Ph-E clusters we detect $37819$ and $47,927$ subhalos, respectively, with masses above \hbox{$10^8$\hmsol}. Of these, $28190$ and $36342$ survive at least $2$ snapshots after accretion, respectively (similar survival times as in the FORS simulation). This ensures that using two cluster resimulations provides similar statistics in the number of subhalos between both sets of simulations.}

{ In our analysis we will concentrate mainly on the FORS simulation, and will only show results
from the Phoenix clusters when there are trends in the statistics associated with higher host halo mass. }
We find that the result that depends most on halo mass is the lifetime of subhaloes (cf. Section $3.3$).

\subsection{Filaments}
In the FORS simulation we detect the filamentary structure around each of the $162$ host halos every $10$ snapshots ($\sim300$Myrs) since $z=1$. We use the Disperse code \citep{disperse} on the particle data within $1\hmpc$ from each halo center. 
Disperse is run with a persistence threshold of $10\sigma$ ensuring only strong filaments, and 
we also remove any filament not passing through halo center. 
A more detailed description of the filament detection method and the selected persistence threshold can be found in the Appendix.

{ In the case of the Phoenix clusters we identify filaments in $42$ snapshots since $z=2$, using the same procedure.}

For each subhalo we compute its distance to the closest filament at accretion time. 
This allows us to define two samples. 
A first selection corresponds to subhalos closer than $40\hkpc$ to a filament at accretion time.  These comprise the Filament sample. 
Second, subhalos more distant than $40\hkpc$ to a filament at accretion time will be the initial candidates for the No Filament sample.

Figure \ref{fig1} shows the mass function for the Filament and candidate No Filament samples in the FORS simulation.
In further analysis, in order to remove any effect from subhalo mass on the properties of the Filament and No Filament samples, we cut the No Filament sample so as to reproduce the same mass function as the one of subhalos in the Filament sample (dashed blue line).  From this point on, this sample will be denoted simply as the No Filament, or Control sample.

For each host we compute the fraction of subhalos accreted from filaments relative to the total subhalo population accreted since $z=1$, and also the fraction of mass contained in such subhalos.
In the bottom-left panel we show corresponding distributions for the number fraction(solid line) and the mass fraction(dashed line) of subhalos accreted from filaments.
In most host halos, less than $40\%$ of their subhalos were accreted from filaments, with an average value close to $20\%$. In terms of mass, the distribution is wider and hosts can have up to $80\%$ of the total accreted mass in subhalos coming from filaments, with an average of $\sim 40\%$.
This can be seen in the high mass-end of the figure where both mass functions (Filaments and control sample) are closer to each other and even overlap at $M\sim10^{10}$\hmsol. 

In the { Phoenix simulations $27-30\%$ of the subhalos were accreted from filaments}, and they account for $\sim 50\%$ of the total subhalo mass accreted since $z=1$.

The question of what is a filament is still not fully settled, and as a result there are several definitions and available methods for filament detection.  There is an ongoing discussion about which method/definition is better, but we do not attempt to enter into this discussion in this paper.
We are aware that using a different filament definition may lead to different results regarding the comparison of the Filament and No Filament samples.  Consequently, we have tested the trends and conclusions along the paper resulting from changing the parameters used to define a filament and find no significant differences.  This may be due to the use of only the strongest filaments in this work, which allows us to remain in the safe zone of convergence among several methods including visual inspection. Lower thresholds that would include weaker filaments would only reduce the strength of the differences between Filament and No Filament samples but would not change the trends and neither would it change the conclusions.

We choose the Disperse method, because it is suitable for the particle distributions around our host halos. It uses a simple definition for filaments based on a single parameter, the persistence threshold, and this method has been used in several simulations and observational data-sets.

For future comparisons we report our average number of $3$ filaments connected to each host, which can be used to tailor samples of filaments obtained using a different filament detection methods with the aim to compare to our findings.  For example, one could choose the N strongest filaments such that the average of filaments connected to each host is the same as ours.

The two parameters governing our filament sample are:
Persistence threshold: this defines the strength of the filament arcs and is measured as the number of standard deviations, $\sigma$, the density of the filament lies above what would be expected for a random distribution. We choose $10\sigma$ to ensure no spurious detection and in order to avoid any resolution effects (see the Appendix for details). Higher threshold values do not change the results of this paper.  However, a threshold below $\sim 6\sigma$ tends to add too many spurious filaments with the effect of diluting the strength of the differences and trends found in our paper.

Distance to filament($D_{FIL}$): Disperse method only resolve the skeleton of the filamentary structure so we must define a filament radius or thickness to identify the volume where subhalo are assigned as filament members. We adopt $40\hkpc$ because this is the typical radius of detected filaments, and this is large enough for statistics purposes where $\sim 20\%$ subhalos are found in filaments. Larger radius increase statistics but decrease the strength of the trends found in this paper, and smaller radius boost the strength of these trends but degrade the statistics.
A more detailed explanation of the effect of this parameter can be found in $3.4$.

\section{Results}
\subsection{Subhalo Orbits}

We track the orbits of all subhalos since accretion time and we define several sub-samples based on their fate:

\begin{itemize}
\item Still Alive: Subhalos still orbiting host at \mbox{$z=0$}.
\item Merged: Subhalos merged with the host before \mbox{$z=0$}.
\item Destroyed: Subhalos who disappear inside the virial radius of their host, but do not merge into any other halo. Most of them are associated with subhalos losing mass due dynamical frictions who are reaching mass detection limit around \hbox{$10^7$\hmsol}. Destroyed halos add up to the host halo mass, but did not reach host center as a substructure, they convert into a diffuse component in the host.
\item Merged Other: Subhalos who end up merged into other subhalo. Most of them are subhalo pairs or clumps at accretion time. A detailed explanation can be found in section $3.8.$
\end{itemize}

\begin{figure}[bth]
\begin{center}
\includegraphics[width=.93\linewidth,angle=0]{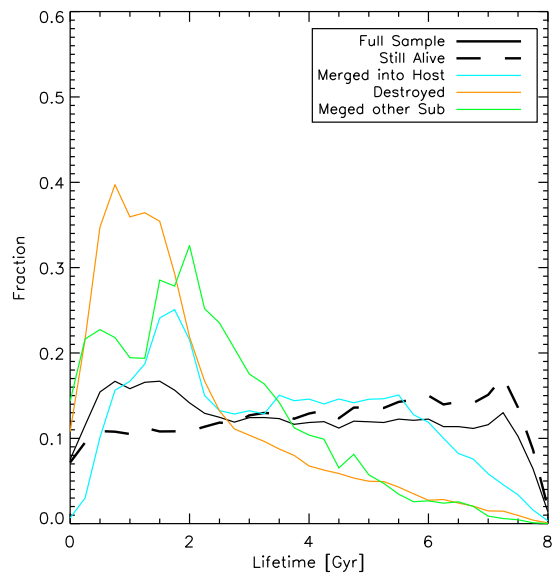}
\caption{
\label{fig2a} Lifetime distribution of subhalos, separated by their final fate. The distributions are truncated at $8$ Gyrs since we consider only subhalos accreted after $z=1$.}
\end{center}
\end{figure}

We define the subhalo lifetime as the time elapsed since accretion into the host and the time they end up destroyed, merged or until $z=0$ if they are still alive.
In the case of subhalos still alive, we remark that the lifetime is the look-back time since accretion, and that a comparison with lifetimes of already dead subhalos(merged/destroyed) should be made carefully.
Figure \ref{fig2a} shows the distribution of lifetimes for the different fates of subhalos. 
The full sample shows a nearly flat distribution truncated at $8$ Gyrs which reflects that we only consider subhalos accreted after $z=1$.
Subhalos still alive show a distribution which is also quite flat, but tends to be older and favor more massive subhalos than the full sample. 
Merged samples show shorter lifetimes and are associated to lower mass subhalos.
Destroyed subhalos show even shorter lifetimes and are associated to even lower masses, in particular low mass subhalos reaching down to the
subhalo mass detection limit due to dynamical friction.

We explore first if the different sub-samples, Filament or Control, suffer different fates.
Table \ref{table1} shows the fraction of subhalos depending with different fates, for subhalos accreted through filaments or otherwise.
Our findings indicate that $\sim70\%$ of subhalos accreted in the last $8$ Gyrs are still alive, $\sim8\%$ merged with the host, $\sim20\%$ were destroyed inside the halo, and $\sim2\%$ merged with another substructure.
In the Phoenix simulations the resulting fractions are similar. 

\begin{table}[ht]
\centering
\begin{tabular}{|l|c|c|}
\hline
Subhalo Fate & Filament & No Filament\\
\hline
Still Alive  &  $0.678 \pm 0.008$ & $0.729 \pm 0.007$\\
Merged       &  $0.080 \pm 0.003$ & $0.073 \pm 0.003$\\
Destroyed    &  $0.214 \pm 0.004$ & $0.175 \pm 0.005$\\
Merged Other &  $0.028 \pm 0.002$ & $0.023 \pm 0.002$\\
\hline
\end{tabular}
\caption{\label{table1} Fraction of subhalos with different final fates for the case when they were accreted through filaments or from the field. Subhalos accreted from filaments are less likely to survive until present time, they are more likely to merge the host, merge with other subhalo and also are more likely to just be destroyed.
 }
\end{table}

Subhalos accreted from filaments tend to have shorter lifetimes and are less likely to be still alive at $z=0$. This result is significant at the $6\sigma$ level. In addition, they are more likely to merge with the host or be destroyed.
This is expected in the scenario that satellites coming from filament follow more radial orbits and reach the halo center straight away in shorter timescales. However, we sill show in section $3.2$ and $3.7$ that this depends on the filament strength and host halo mass. 
In the case of subhalos merged with another subhalo, this is more likely to happen in filaments mainly because clumps or groups of subhalos are more common in large-scale filaments than in the field, and therefore are more likely to have been part of a group of subhalos before accretion. We will study this scenario in more detail in section $3.8$.

\subsection{Influence of filamentary accretion on subhalo lifetimes}

\begin{figure}[!h]
\begin{center}
\includegraphics[width=.95\linewidth,angle=0]{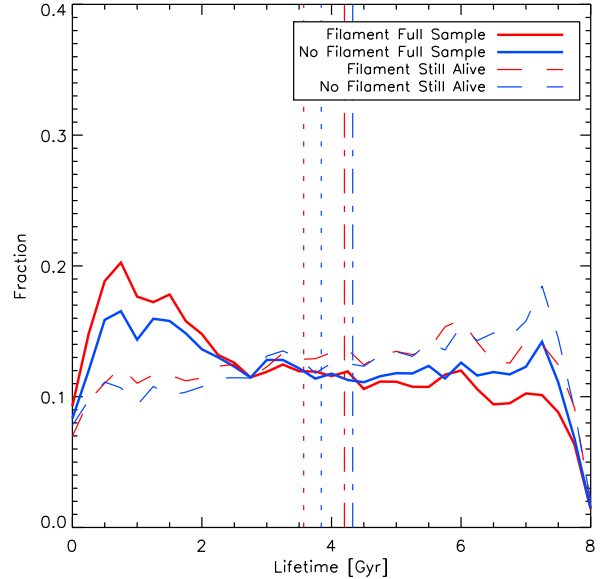}
\caption{ Lifetime distribution for subhalos accreted through filament or from the field. For the full  (solid lines) and Still Alive samples (dashed lines), subhalos accreted from filaments show shorter lifetimes.
\label{fig2b} }
\end{center}
\end{figure}

We explore the relation between filamentary accretion and subhalo lifetime for different subhalo samples.
Figure \ref{fig2b} shows lifetimes since accretion for subhalos accreted from filaments and in the control sample. We find mean lifetimes of $3.57\pm0.02$ Gyr for the Filament sample and $3.86\pm0.03$ Gyr for the control sample, this is, subhaloes arriving through filaments show a $\sim8\%$ shorter lifetime at a $9\sigma$ significance. { The average values and fractional differences between filament and no filament samples are summarized in table \ref{table2}.} 
In the case of the Still Alive sample the lifetimes are $4.20\pm0.03$ Gyrs for the Filament sample and $4.35\pm0.03$ Gyrs for the control sample; only $3-4\%$ shorter lifetimes for subhalos arriving via filaments.
For the Merged sample lifetimes are $3.35\pm0.03$ Gyrs and $3.62\pm0.05$ Gyrs respectively, $8\%$ shorter for subhalos arriving through filaments.  For the Destroyed sample, lifetimes are $1.85\pm0.02$ Gyrs and $2.10\pm0.03$ Gyrs respectively, that is, $12\%$ shorter lifetime for the filament case.

The result of larger lifetime difference for merged and destroyed samples compared to the still alive sample should be taken with caution since in these cases, the lifetime definition does not account for the total survival time of the halo, as it represents the look-back time since accretion and does not take into account the total time it will survive.

We explore if the shorter lifetimes of subhalos accreted from filaments depend on the properties of their host halos.
First, we select host halos by mass growth ratio $M/R$ already defined in Section $2.2$.  There is no clear difference in the subhalo lifetimes and filamentary accretion which indicates the speed of growth of the hosts is not an important factor.
Second, we explore the dependence on the fraction of subhalos accreted from filaments using the number and mass fractions defined in Section $2.4$.  We find only a small signature that lifetime difference between Filament and control sample is slightly larger for hosts having a larger filament mass fraction. This difference is probably produced by the larger fraction of mass accreted from filaments in lower mass hosts.  In the next subsection we explore the effect of host mass.

\subsection{Dependence on host halo mass}

To study the effect of host halo mass in the lifetimes of subhalos accreted from filaments or control samples, we make use of FORS and Phoenix simulations. With the FORS simulation we can explore low- and mid-range host masses; in Phoenix clusters we reach the high mass end.

\begin{figure*}[!htb]
\begin{center}
\includegraphics[width=.32\linewidth,angle=0]{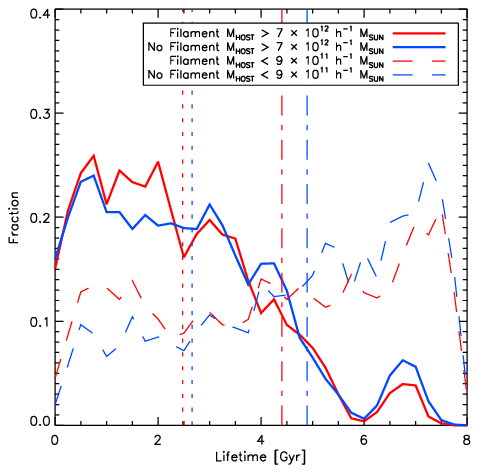}
\includegraphics[width=.32\linewidth,angle=0]{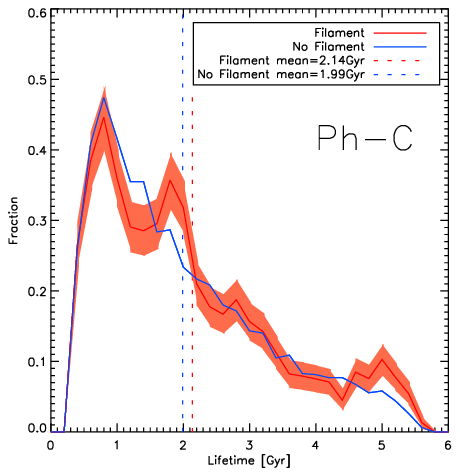}
\includegraphics[width=.32\linewidth,angle=0]{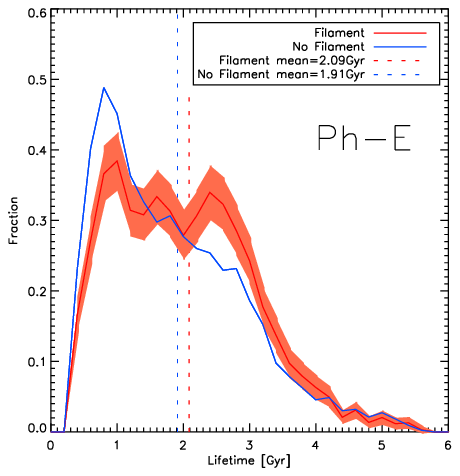}
\caption{
\label{figph} Distribution of lifetimes for subhalos accreted through filaments or away from them, for different host halo masses. Left panel: Hosts are selected by mass at subhalo accretion for the low (dashed) and high mass range(solid) in the FORS simulation. 
{ Middle and right panels: results for the Phoenix C and E clusters. Vertical lines show mean lifetimes. The lifetime difference between the filament and control samples depends on host halo mass. In lower mass hosts lifetimes are shorter in Filament samples, but surprisingly this relation weakens at intermediate masses, and even reverses in the massive Phoenix clusters.}  
The latter can be understood
recalling that more massive host halos are connected to more numerous, stronger filaments, which are thicker and with higher velocity coherence; subhalos accreted through filaments, even if they have more radial orbits, are shielded from dynamical friction in their first infall by this strong inter-filament coherent flow.}
\end{center}
\end{figure*}

Figure \ref{figph} shows the lifetime distributions for subhalos accreted from filaments and for the control sample, for different host masses. The left panel shows results for the $20\%$ percentiles of low ($M_{HOST}<9 \times 10^{11}$\hmsol) and mid ($7 \times 10^{12}<M_{HOST}<10^{13}$\hmsol) mass ranges in the FORS simulation. 
{ The middle and right panels shows results for the Phoenix clusters($M_{200} \sim 6 \times 10^{14}$\hmsol).}

First, subhalos in higher mass host halos have shorter lifetimes in general, this is because they are subjects of more dynamical friction in a denser environment.
Second, the difference in lifetime between subhalos accreted from filaments and the control sample is slightly larger for lower mass halos, { specifically, $10\%$ shorter lifetimes for low mass hosts, and $8\%$ shorter lifetimes for intermediate mass hosts. In lower mass hosts the mean lifetimes are $\sim 0.5$ Gyr shorter for the Filament sample, and in intermediate hosts it is $\sim 0.2$ Gyr shorter.
}
{ In the high mass hosts (Phoenix clusters) subhalos accreted from filaments have longer lifetimes ($\sim 0.2$ Gyr) when compared to the control sample.}
This reveals a trend that in higher mass hosts the lifetime difference for subhalos accreted from filaments becomes weaker and even reverses for cluster sized halos.

This behavior is somewhat expected as more massive host halos are connected to, denser and stronger filaments, which are thicker ($0.5-1 \hmpc$ diameter in cluster sized halos) and with higher velocity coherence. 
Therefore, the subhalos accreted through filaments, which { appear to show slightly more radial orbits}, are shielded from dynamical friction in their first infall by the dense and coherent flow of the filament they are falling through.
{ In section $3.6$ and Appendix B we explore in more detail the orbital parameters of subhalos accreted through filaments.}

Given the larger spacing between simulation snapshots in the Phoenix simulations, we will restrict most of the remainder of the analysis, which will now focus on orbital parameters, to the high resolution simulation, i.e. to halos with masses of groups and lower.

\subsection{Robustness to sample definition}

\begin{figure*}
\begin{center}
\includegraphics[width=.95\linewidth,angle=0]{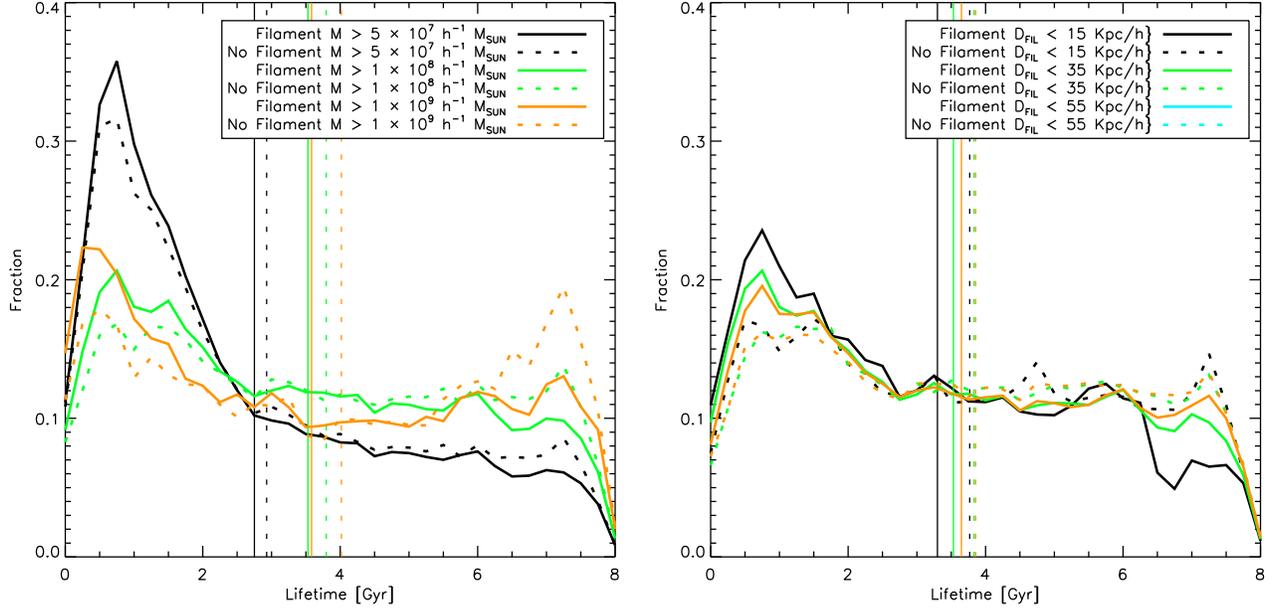}
\caption{
\label{fig3} Lifetime distributions for Filament and No Filament samples, for different sample definition parameters.  Left panel: subhalo mass. Right panel: Filament distance criterion. Vertical lines indicate mean values for each corresponding sample. Differences in the lifetime for Filament and control sample are stronger for higher subhalo masses and lower distances to filaments.}
\end{center}
\end{figure*}

We now explore the dependence on subhalo mass and distance filament skeleton used originally to define the Filament and No Filament samples. Figure \ref{fig3} shows the distributions of lifetimes for different subhalo masses, in the left panel, and for different distances to the filament in the right panel.

\begin{figure*}[!htb]
\setlength{\abovecaptionskip}{15pt plus 3pt minus 2pt}
\setlength{\belowcaptionskip}{15pt}
\begin{center}
\includegraphics[width=.98\linewidth,angle=0]{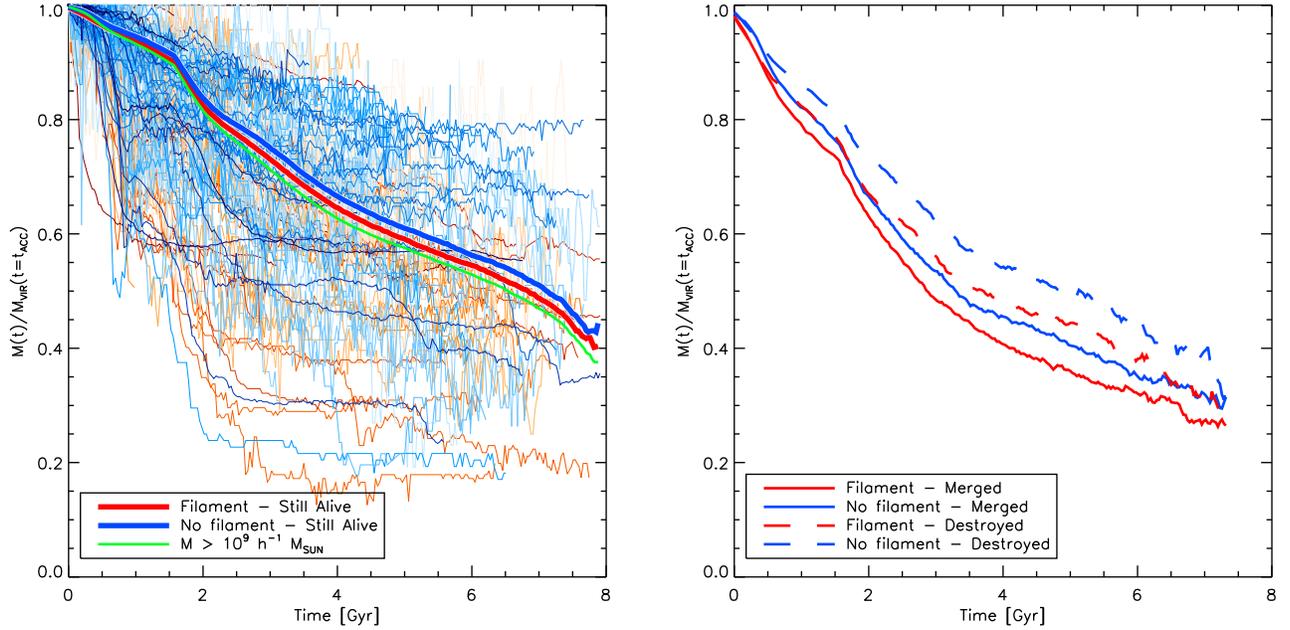}
\caption{
\label{fig4} Average fractional mass loss for subhalos in Filament and control samples as function of time elapsed since accretion. Left panel: Still Alive subhalos (average value shown as thick lines), secondary thin lines show a few individual subhalos in the Filament(red) and No Filament(blue) samples; darker colors indicate more massive subhalos. The green line shows the average fractional mass loss for high mass subhalos.
Right panel: subhalos in the Merged and Destroyed sub-samples of the Filament and control samples, where the differences due to filamentary accretion are larger.
Subhalos in Filament sample lose mass more rapidly in agreement with shorter lifetimes. More massive subhalos lose mass more rapidly.}
\end{center}
\end{figure*}

The left panel of the figure shows that lower mass subhalos have shorter lifetimes as expected because they lose mass due to dynamical friction and fall below the subhalo detection limit earlier. 
The lifetime differences between Filament and control samples for low mass subhalos is of shorter timescales by $6\%$. For high mass subhalos timescales are $12\%$ shorter.
The right panel shows an important increase in the lifetime difference for shorter values in the criterion taking into account distance to filament. For $D_{FIL} < 15\hkpc$ lifetimes are $14\%$ shorter, and for $D_{FIL} < 55\hkpc$ these are only $5\%$ shorter. This is a clear signature that the methodology to detect the filaments adopted in this paper is indeed selecting filamentary regions where subhalo accretion is different, with { a tendency toward more radial orbits around} the halo centers, which is reflected in shorter subhalo lifetimes.

\subsection{Mass loss after accretion}

We explore the mass loss of subhalos since accretion time.  Dynamical friction studies show that subhalo mass loss and lifetime depend on how deep the subhalos penetrate the host halos, and this is related to orbit eccentricity, and also subhalo-to-host mass ratio \citep{2008MNRAS.383...93B}.

Figure \ref{fig4} shows the average subhalo mass as a function of time since accretion, with mass normalized by the mass at the moment of accretion. 

The left panel  shows the average fractional mass loss for the Still Alive sample (thick lines).  Thin lines show the individual fractional mass evolution for $1\%$ of the subhalos for the Filament(red) and No Filament(blue) samples, with darker colors corresponding to more massive subhalos and lighter colors to less massive ones. 
Average values are computed with an appropriate weighting at every time step since each subhalo has different lifetimes which makes them contribute to fragments of the curve showing the average.
The solid thick lines shows the average values; as  can be seen,  subhalos accreted from filaments have higher mass loss in concordance with their shorter lifetimes.
Green shows the average fractional mass for massive subhalos which lose mass faster than lower mass subhalos, as expected from dynamical friction effects \citep{2008MNRAS.383...93B}.

\begin{figure*}[!ht]
\setlength{\abovecaptionskip}{10pt plus 3pt minus 2pt}
\setlength{\belowcaptionskip}{12pt}
\begin{center}
\includegraphics[width=.95\linewidth,angle=0]{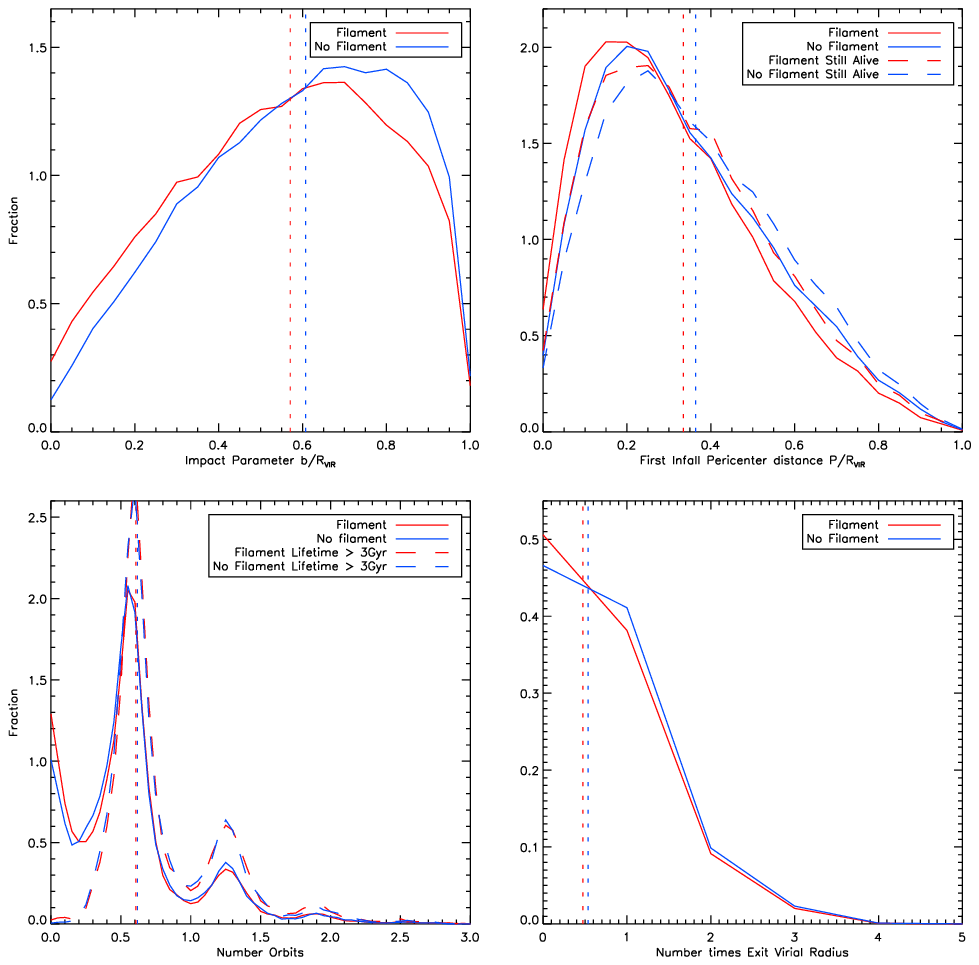}
\caption{
\label{fig5} Orbital Parameters for subhalos in Filament and No Filament samples. Top left: Impact parameter computed using subhalo velocity vector at accretion. Top right: Pericenter distance in the first orbit. Bottom left: Number of orbits. Bottom Right: Number of times the subhalo escapes the virial radius after accretion. 
{ Average values for each sample are shown as vertical dotted lines.}
Subhalos accreted from filaments have { a tendency to show } more radial orbits at infall, reach closer pericenter distances, have smaller number of orbits, and are less likely to escape the virial radius, even after taking into account their  { higher ellipticity} orbits.}
\end{center}
\end{figure*}

The right panel shows the fractional mass loss for the Merged and Destroyed samples. We found that mass loss difference between Filament and control sample for Merged and Destroyed subhalos are larger than for the Still alive ones, in agreement with larger differences in lifetimes for corresponding samples aswell.
In addition, the mass loss for the Merged sample is larger than for the Destroyed sample; naively  we would expect the opposite, that the Destroyed sample end mass
 should even reach values close to zero, which is not the case.  This reflects that the Destroyed sample is dominated by low mass subhalos which disappear due to resolution problems rather than reaching a small fraction of their initial accretion mass.
Notice that these low mass subhalos do not affect lifetimes differences found in this paper, since we can see same trends in Still Alive samples or using larger subhalo masses(figure \ref{fig3}).

\subsection{Orbital parameters}

We measure several orbital parameters for each accreted subhalo and show the results in
Figure \ref{fig5}, { and average values are shown in table \ref{table2}}

Impact parameter: the top left panel  shows the distributions of impact parameters computed using the subhalo velocity vector at accretion. Subhalos in the Filament sample show smaller impact parameters indicating that subhalos accreted through filaments { with a tendency for} more radial orbits.

Pericenter distance: the top right panel  shows the distributions of pericenter distances in the first orbit.
Subhalos in the Filament sample show smaller pericenter distances, in agreement with smaller impact parameter and { a tendency for} more radial orbits. Notice that this pericenter distance is computed only for subhalos that make at least half an orbit in order to ensure the pericenter distance is the minimum radial distance along the trajectory and not just the extreme of one small orbit segment, for the case of very short lived subhalos.

Number of Orbits: the bottom left panel  shows the distribution of fractional number of orbits.
Subhalos in the Filament { sample make less orbits but it is barely significant, however this slight trend is} in agreement with their shorter lifetimes.
We can see the resonances at $0.6,1.2,1.8$ orbits due to the likelihood that for elliptic orbits they merge or get destroyed slightly after passing pericentric points which on average are located in such positions. 
We select a sample of long lived subhalos with lifetimes $>3$ Gyr to check that there is no difference in the number of orbits due to filamentary accretion, and find resonance peaks in the same positions.

Number of times reaching outside the virial radius: the  bottom right panel  shows the distribution of the number of times subhalos escape the virial radius since accretion. We find a larger number for the No Filament samples. 
Even though we expect a larger number of escapees for more radial and eccentric orbits, such as found in filaments, we find the opposite effect.  This is due to subhalos in the Filament sample making less orbits which makes it less likely for them to survive enough to escape beyond the virial radius.

\subsection{Lifetime and filament strength}
\label{ssec:lifetime}

\begin{figure}[!h]
\begin{center}
\includegraphics[width=.97\linewidth,angle=0]{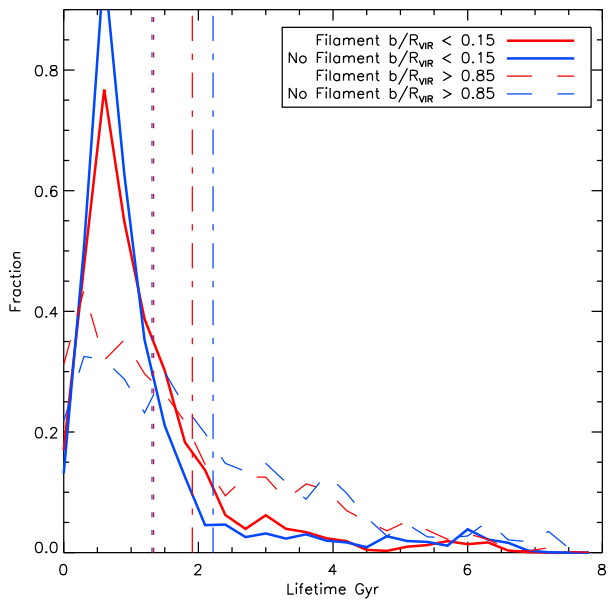}
\caption{
\label{fig6} Lifetime distribution for Destroyed subhalos in Filament and control samples having low(solid lines) and high(dashed lines) impact parameter at accretion.
Subhalos in filaments with low impact parameters can be associated with strong filaments with { slightly stronger} radial accretion and higher velocity coherence.  For high impact parameter, they can be associated with diffuse filaments.
For subhalos accreted through stronger filaments the lifetime difference between Filament and control sample is strongly reduced.}
\end{center}
\end{figure}

In this section we explore the how the lifetime difference between Filament and control samples depends on filament strength.  We define as strong filaments the ones that are
more dense and thick, which in general feed their connected hosts with a more coherent matter flow.

We infer that low impact parameter subhalos in the Filament sample are aligned with the filament direction at accretion, so it is more likely they come from strong filaments with a more coherent radial flow, and in an analogous way, high impact parameter subhalos in the Filament sample are misaligned with the filament, so it is more likely they come from weaker and more diffuse filaments.
Figure \ref{fig6} explores this relation for Destroyed samples where subhalos are more likely to die faster,
so as to isolate what happens in the first infall through a filament.
We find that for the low impact parameter sample, the difference in lifetime totally disappears, and for high impact parameter it reverts to $\sim 13\%$ shorter lifetimes for the Filament sample. 
This effect is also present in the Still Alive and Merged samples.

If we join this result with the dependence on host halo mass, we can infer that the lifetime difference dissipates and even reverses for larger host halos connected to stronger filaments, where the filaments have stronger velocity coherence and are thick enough to shield the first infall of subhalos from dynamical friction. 
\begin{figure}
\begin{center}
\includegraphics[width=.97\linewidth,angle=0]{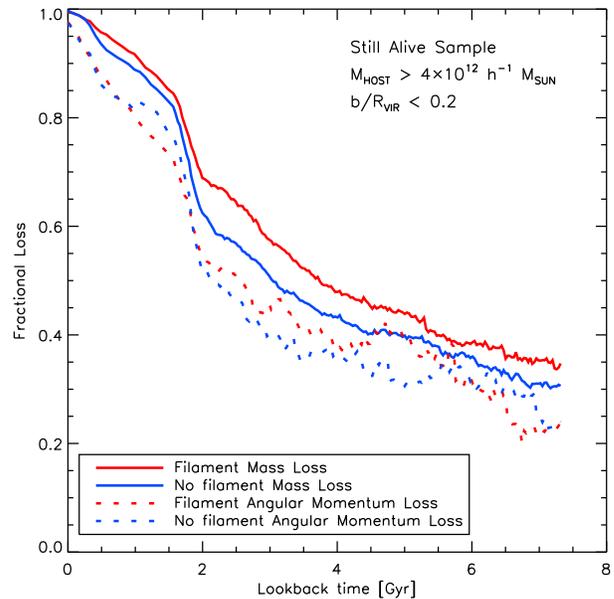}
\caption{
\label{figang} Evolution of the average fractional mass loss(solid lines) and angular momentum loss(dashed lines) since accretion, for Still Alive subhalos accreted from stronger and more coherent filaments selected in higher mass hosts. Compared to Figure \ref{fig4}, the relation reverses and subhalos accreted from strong filaments lose mass at a slower rate than the control sample. The fractional angular momentum loss seems to be important in the first Gyr after accretion, but it follows the mass loss trend at later times.}
\end{center}
\end{figure}

\begin{figure*}
\begin{center}
\includegraphics[width=.95\linewidth,angle=0]{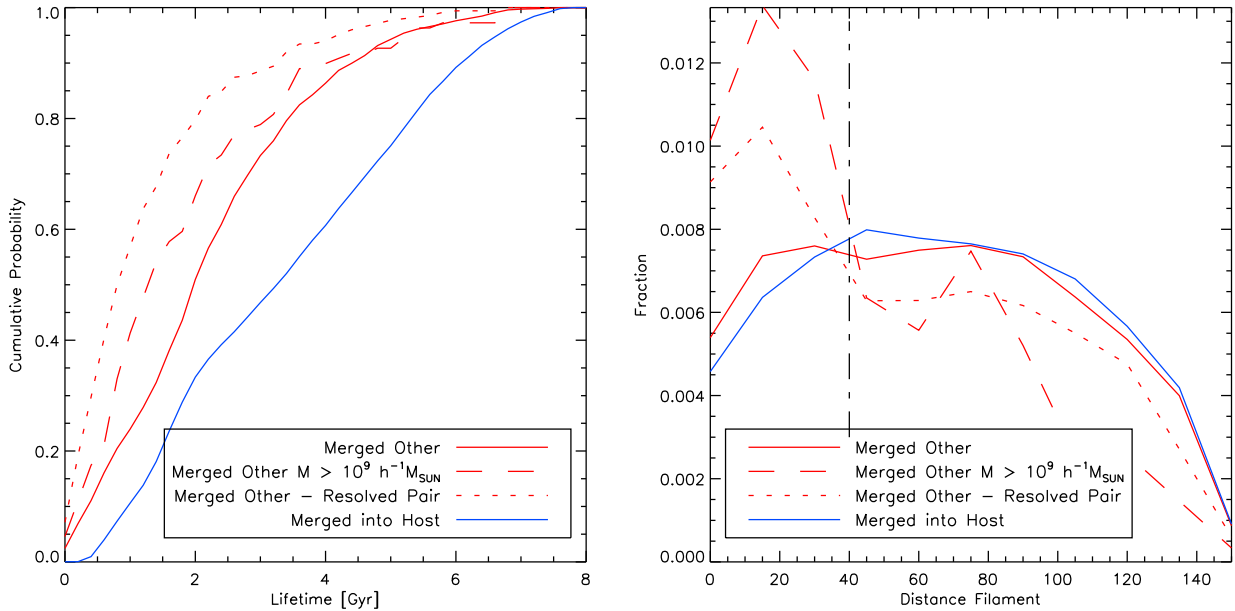}
\caption{
\label{fig7} Left Panel: Cumulative probability distribution of lifetimes for subhalo pairs/clumps and subhalos merged with the host (see the key). Subhalo pairs or groups (such as for example the Magellanic Clouds or other MW satellite associations) have shorter lifetimes, on average. Right panel: distribution of Distance to filament for the same samples. There is a clear signature that subhalo pairs/clumps are more likely to be accreted through filaments when compared to single subhalos.}
\end{center}
\end{figure*}

We test this hypothesis using the most massive hosts in the FORS simulation with $M > 4\times10^{12}$\hmsol at $z=0$, selecting subhalos still alive with low impact parameter $b/R_{VIR}<0.2$. Figure \ref{figang} shows the average fractional mass (solid lines) and angular momentum (dotted lines) loss for Filament and control samples. Here we confirm the effect measured in the Phoenix simulations where subhalos from filaments lose mass and angular momentum at a slower rate than those in the control sample, indicating that strong filaments shield subhalos from dynamical friction.
The angular momentum evolution is quite steep the first Gyr, but then it tends to follow the slope of the mass evolution.

\subsection{Satellite pairs/clumps and the Magellanic Clouds}

We focus our attention to the $\sim2\%$ subhalos which end up merged with another subhalo.
 The Merge Other sample is defined as all subhalos which suddenly increase their mass by a factor $\geq2$ between two consecutive snapshots after accretion, and survive for at least $2$ more snapshots with a similar or even larger mass. 
We use this more general definition rather than Rockstar halo finder merge conditions which requires two well resolved subhalos and may skip some transient effects or disrupted subhalos. 

We find that most of these subhalos increase their mass due an important accretion/merge event, however in some cases the other merging member may be a non resolved accreted subhalo.

We define an additional subhalo sample of Resolved Pairs, which is a sub-sample of the Merge Other sample but with the additional constraint that the other merging member is a well resolved subhalo. Around one third of the Merge Other sample are Resolved Pairs.

We can associate these sub-samples, in particular the Resolved Pairs sub-sample, as subhalo pairs or associations which end up merged together before they are destroyed by dynamical friction or merged with the host. 
Most of these subhalos are accreted with their merger companion already nearby, probably bound, and it is extremely rare to find one subhalo merging with another subhalo just by a random encounter after accretion.
In \citet{2013ApJ...770...96G} they find a similar incidence of $\sim 2\%$ subhalo pairs in MW-sized halos, and this drops to $\sim1$ in a sample of $30000$ if we look for matching MC analogues.

We found that only $\sim9\%$ of these subhalo pairs survive until $z=0$, and most of them accreted less than a Gyr ago.

Figure \ref{fig7} explores the cumulative probability distribution of their lifetimes in the left panel, where lifetime is defined as the time elapsed since pair/clump accretion until they merge each other.
By definition the time when they merge is the time when the major interaction begins, the transient time until the merging process stabilizes may last up to a couple of snapshots $\leq 100 Myr$.

We find shorter lifetimes compared with subhalos merged with the host (blue line), and lifetimes become shorter for larger subhalo masses, and for the Resolved Pairs sub-sample.
 For subhalos larger than $10^{9}$\hmsol we find a $50\%$ chance their lifetimes are shorter than $1.4$Gyr and a $90\%$ chance their lifetimes are $<4$Gyr.  For Resolved Pairs the $50\%$ and $90\%$ chances drop to $0.8$ Gyr and $3.4$ Gyr respectively. 
 If we associated these subhalos with the Magellanic Clouds, we would infer they entered the MW virial radius very recently and probably are in the first infall, or at least have traversed less than a full orbit, otherwise they should be already merged.
This supports the first infall scenario proposed by \citet{2007ApJ...668..949B} using HST proper motions of the MCs, and is in agreement with the results of \citet{2015arXiv150703594P}.

In the right panel of the figure we show the distribution of distance to filament for the same sub-samples compared with subhalos merging into the host. We find that all samples related to subhalo pairs/clumps are more likely to be closer to filaments at accretion. For subhalos merged into host, $19\%$ of them were accreted from filaments($d < 40\hkpc$), for the Merged Other sample this fraction raises to $21\%$, for Resolved Pairs sub-sample it increases to $29\%$, and for Merge other subhalos larger than $10^{9}$\hmsol the fraction is $39\%$.
This summary along with the result shown in the last row of Table $1$, indicate that filamentary accretion boosts the formation of subhalo associations.  As a result, host halos having more filamentary accretion along their formation history are expected to have more subhalo pairs such as the MCs.

In the Local Group, the overabundance of subhalo pairs in the MW and M31 compared with \LCDM halos of similar mass \citep{2013MNRAS.431L..73F} may be explained in the scenario where the process of filamentary accretion was dominant in the formation of the main galaxies of the LG. We do not know the assembly history of the LG, but at the present time the MW and M31 are located in a filamentary structure \citep{2013AJ....146...69C}, and from numerical simulations it is believed that MW-M31 pairs are very likely to be located and aligned with large scale filaments \citep{2015ApJ...799...45F}.

\section{Discussion and conclusions}

We studied the lifetime of satellites, and how this depends on whether the satellite arrived via a large-scale filament, or from the field. 
We also studied how this changes with the host halo mass.
In the case of low mass halos, subhalos accreted through filaments show $\sim 10\%$ shorter lifetimes compared to subhalos accreted from random directions. This is expected since subhalos accreted through filaments have { a preference for} more radial orbits and are more likely to reach halo central regions earlier, suffering stronger dynamical friction and merging with the host.
In higher mass haloes we notice that this effect dissipates, and when we select the strongest filaments where there is higher velocity coherence along the filament (in the radial direction to the host center) we notice also that this effect dissipates. This may be explained because larger mass hosts are connected to stronger filaments with higher velocity coherence and density, where even though orbits are { slightly} more radial, the filament is thick enough to shield the subhalo at least in its first infall, reducing dynamical friction effect because the subhalo follows the filament's coherent flow.
We tested this hypothesis using both simulations suites FORS and Phoenix.  In the former we selected a sub-sample comprising the most massive hosts with low impact parameter subhalos. In the Phoenix simulations we simply have two clusters of $\sim 6 \times 10^{14}$\hmsol.  In both cases, the difference in lifetimes reverses, and subhalos accreted through strong filaments present larger lifetimes than subhalos accreted from random directions.

This shielding from dynamical friction produced in filaments during the first infall is also important to understand the segregation of subhalos by their masses at accretion. \citet{2015arXiv151001586V} demonstrate that part of this segregation is already imprinted in the infall conditions. For massive subhaloes it is subsequently boosted by dynamical friction, but only during their first radial orbit. 
 
The trends with filamentary accretion found for the lifetimes are not strong enough to influence general halo statistics or cosmology, but for precision models of dynamical friction and tidal stripping this should be taken into account as it would change the properties of the satellite population produced in SAMs or HOD models.  It can also help to understand better assembly bias or other clustering systematics for future surveys, and can be a key signature to take into account when studying the planes of satellites recently found in the Local Group.

Subhalo pairs or groups, similar in nature to the Magellanic Clouds or other MW satellite associations, have shorter lifetimes. 
For Merge Other subhalos (pairs of subhalos that merge each other) larger than $10^{9}$\hmsol we found a $50\%$ chance their lifetimes are shorter than $1.4$Gyr, and a $90\%$ chance their lifetimes are shorter than $4$Gyrs. 
For the sub-sample of Resolved Pairs these lifetime percentiles are even shorter.
By associating these subhalos with the Magellanic Clouds one can infer they have entered the MW virial radius very recently and probably are in their first infall, or are likely still within their first full orbit, otherwise they should have already merged.
We also found that filamentary accretion boosts the accretion of subhalo pairs/groups, and this may be an important clue for understanding the overabundance of subhalo pairs in the Local Group.

\acknowledgments
We have greatly benefited from discussions and revisions of the near to final draft by Chervin Laporte and Simon White.
REG was supported by Proyecto Financiamiento Basal PFB-06 'Centro de
Astronomia y Tecnologias Afines' and Proyecto Comit\'e Mixto ESO
3312-013-82.
NP acknowledges support from Fondecyt Regular 1150300, and thanks the hospitality of the Max Planck
Institute for Astrophysics where he spent his 2013-2014 Sabbatical year.
The Geryon cluster at the Centro de Astro-Ingenieria UC was
extensively used for the calculations performed in this paper. The
Anillo ACT-86, FONDEQUIP AIC-57, and QUIMAL 130008 provided funding
for several improvements to the Geryon cluster.

\appendix

\section{Filament detection sensitivity}

We identify filaments in this paper using Disperse code \citep{disperse}.
This code uses Morse theory to identify critical points in a point distribution, then it connects these critical points, in particular maximum and saddle points to form arcs, then a set of connected arcs form a skeleton which is the central tracer of a filamentary structure.
In the context of Morse Theory, a persistence pair is a pair of critical points with a critical index difference of 1(i.e. Maxima and Saddle point), and the persistence is a positive value to represent the importance of the persistence pair in comparison with the noise. 
Higher persistence threshold to identify filaments mean stronger filaments with a higher contrast among a random distribution of points. 
This threshold can be measured in terms on the number of sigmas, where the persistence pair has a probability of at least n-sigma to not appear in a random field.   
Disperse ran with a persistence threshold of $10\sigma$ ensuring only strong filaments not produced by random.

\begin{figure}
\begin{center}
\includegraphics[width=.95\linewidth,angle=0]{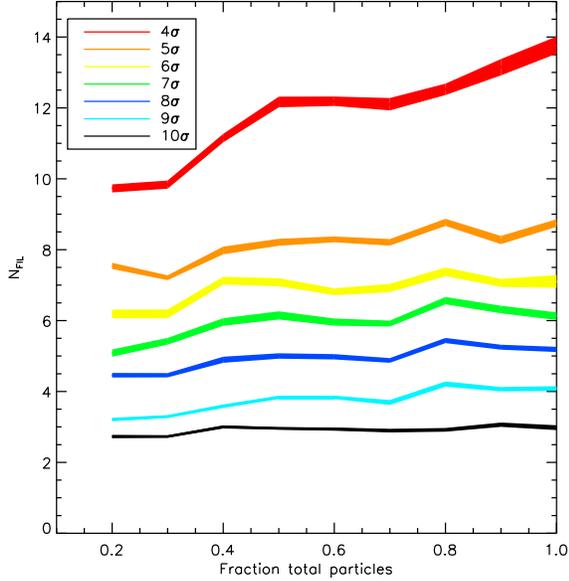}
\caption{ 
\label{figrestest} Resolution test: average number of filaments dependence on the fraction of simulation particles used.
We use $50$ host halos and we run Disperse code on sub-samples having from $20$ to $100\%$ of the original particles. 
Higher detection threshold results in less filaments, but with higher contrast. 
We can clearly see that filaments detected with persistence threshold of $10\sigma$ are strong enough to be fully recovered even when using only $20\%$ of the original particles.
}
\end{center}
\end{figure}

In addition we test the effect of simulation resolution on the persistence threshold, the idea is check if the number of particles we use to sample the filaments around the host halos is enough to identify consistent filaments and if this is not resolution dependant.
For this purpose we select $50$ halos at $z=0$ and compute their filaments using form $20\%$ to $100\%$ of the total number of sampling particles. Filaments are also computed for several persistence thresholds from $4$ to $10\sigma$. Results are shown in Figure \ref{figrestest}, we have the average number of filaments dependence on the fraction of original particles used, and for different detection thresholds. 
Errors are represented by the thickness of each line. We have that at higher detection thresholds, we identify less filaments as expected, but they are stronger an less dependent on the fraction of particles used. 
At $4\sigma$ we have an average of $10$ filaments per halo, but we miss around $30\%$ of the filaments if we use only the $20\%$ of the particles; However at $10\sigma$ we found an average of $3$ filaments per halo, but this number does not depend on the number of particles used. This demonstrates that we do not suffer resolution effects on the detection of filaments and they are not spurious detections.

After we have the filaments for each halo at a given redshift, we select only the filaments which are connected to a maximum critical point within $0.5$ times the virial radius. We discard filament data beyond two times the virial radius.
The final step is to compute for each subhalo the distance to the closest filament segment.

\section{{ Increased ellipticity of} orbits from filaments}

\begin{figure}
\begin{center}
\includegraphics[width=.95\linewidth,angle=0]{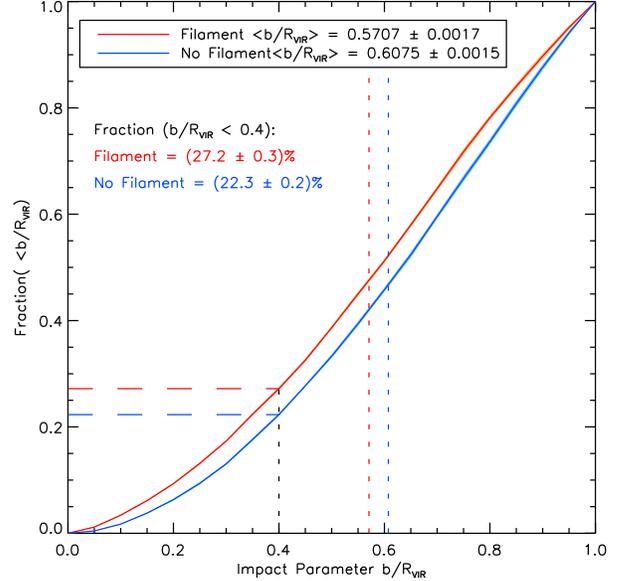}
\caption{
\label{figrefer1}
{ Cumulative distribution of impact parameters for subhalos accreted through filaments or from the field. There is a significant trend that subhalos accreted through filaments have more radial orbits, with a lower impact parameter at accretion.}
}
\end{center}
\end{figure}

{
We explore in more detail the trends found in figure \ref{fig5} where subhalos accreted through filaments have lower impact parameters at accretion and smaller pericentric distances indicating { a tendency for more} radial orbits. In figure \ref{figrefer1} we show the cumulative distribution of impact parameters for subhalos accreted through filaments and for control sample, where  $27.2\pm0.3\%$ of subhalos in the filament sample have impact parameters lower than $0.4$.  This fraction drops to $22.3\pm0.2\%$ in the no filament sample.
}

\section{Statistical significance summary}

{
For the differences in the properties between subhalos accreted from filaments and control sample, we compute average values and associated errors using the jacknife method. Table \ref{table2}  summarizes these measurements making reference to the figures where these results are shown.
}

\begin{table*}[ht]
\centering
\begin{tabular}{|l|l|c|c|c|}
\hline
Figure & Sample & Filament & No Filament & Difference(\%)\\
\hline
4 & Full          &  $3.57 \pm 0.02$ & $3.86 \pm 0.03$ & $7.5$\\
  & Still Alive   &  $4.20 \pm 0.03$ & $4.35 \pm 0.03$ & $3.4$\\
\hline
5 Left & Large $M_{HOST}$   &  $2.47 \pm 0.04$ & $2.67 \pm 0.04$ & $7.5$\\
     & Small $M_{HOST}$   &  $4.39 \pm 0.05$ & $4.89 \pm 0.03$ & $10.2$\\
$\, \, \,$ Middle & Ph-C            &  $2.14 \pm 0.04$ & $1.99 \pm 0.03$ & $-7.5$\\
$\, \, \,$  Right  & Ph-E            &  $2.09 \pm 0.04$ & $1.91 \pm 0.02$ & $-9.4$\\
\hline
6 Left  & Low $M_{SUB}$  &  $2.75 \pm 0.03$ & $2.93 \pm 0.03$ & $6.1$\\
      & Medium $M_{SUB}$  &  $3.53 \pm 0.04$ & $3.80 \pm 0.03$ & $7.1$\\
      & High $M_{SUB}$  &  $3.57 \pm 0.06$ & $4.03 \pm 0.04$ & $11.4$\\

$\, \, \,$ Right & Low $D_{FIL}$  &  $3.30 \pm 0.05$ & $3.83 \pm 0.05$ & $13.8$\\
      & Medium $D_{FIL}$   &  $3.53 \pm 0.04$ & $3.86 \pm 0.03$ & $8.5$\\
      & High $D_{FIL}$  &  $3.65 \pm 0.04$ & $3.85 \pm 0.04$ & $5.2$\\
\hline
8       & $b/R_{VIR}$ & $0.5707 \pm 0.0017$ & $0.6075 \pm 0.0015$ & $6.1$\\
        & $P/R_{VIR}$ & $0.3342 \pm 0.0013$ & $0.3637 \pm 0.0016$ & $8.1$\\
        & Number Orbits & $0.6079 \pm 0.0026$ & $0.6192 \pm 0.0020$ & $1.8$\\
        & Number Exits & $0.4726 \pm 0.0027$ & $0.5371 \pm 0.0031$ & $12.0$\\
\hline
9       & Low $b/R_{VIR}$ &  $1.33 \pm 0.06$ & $1.31 \pm 0.05$ & $-1.5$\\
        & High $b/R_{VIR}$ &  $1.91 \pm 0.05$ & $2.21 \pm 0.04$ & $13.5$\\
\hline
\end{tabular}
\caption{\label{table2}
Average values for the different properties such as lifetime and orbital parameters in the filament and no filament samples, that are shown in the different figures as vertical dotted lines.
 }
\end{table*}

\section{Resolution effects: Time sampling}

\begin{figure}
\begin{center}
\includegraphics[width=.95\linewidth,angle=0]{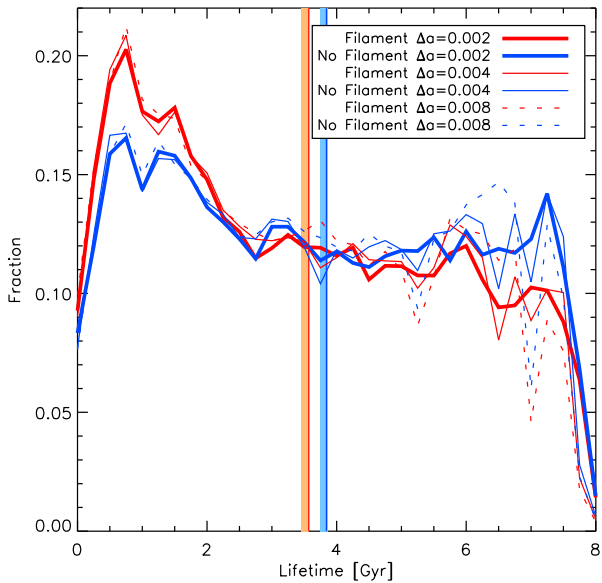}
\caption{
\label{figrefer2}
{
Same lifetime distribution from figure \ref{fig2b}(solid thick lines), but including cases where time resolution is degraded to half(solid thin lines) and to a quarter(dashed lines) of the fiducial timestep. The average values for the full resolution case are shown as vertical solid lines; the degraded cases with errors fall in the respective colored regions.
}
}
\end{center}
\end{figure}

{
We explore the effect of time sampling in the subhalo orbit computation. One of the main features of the FORS simulation is the exceptional high temporal resolution where snapshots are spaced by a scale factor difference $\Delta a=0.002$, where we can track very detailed subhalo orbits. However in most cosmological simulations snapshots are spaced by larger time steps.  Here we test how the measured lifetime differences between filament and no filament samples are affected by the accuracy of the orbit and lifetimes when snapshots are more coarsely spread.

In order to perform this test we compute subhalo orbits using half the snapshots, resulting in a spacing of $\Delta a=0.004$.  We also repeat the procedure using one quarter of the snapshots, resulting in $\Delta a=0.008$.

In figure \ref{figrefer2} we repeat figure \ref{fig2b}, computing lifetime distribution for subhalos accreted through filaments or from the field, but for different temporal resolutions: full resolution(solid thick lines), half resolution(solid thin lines), and quarter resolution (dashed lines). Average values for the full resolution case are shown as vertical solid lines, and average values for the degraded resolution cases including errors lie within in the colored regions.
}

{
We found average lifetimes of $3.52\pm0.03$Gyr and $3.45\pm0.02$Gyr for the filament samples in the half and quarter resolution cases, respectively, and $3.80\pm 0.02$Gyr and $3.74\pm 0.03$Gyr for the no filament samples as well.
We found that lifetimes become systematically slightly shorter as we degrade the resolution, mostly due to dead or merge times becoming shorter in coarser time samplings. However, we found that the fractional difference between filament and no filament samples in all three resolutions remains at $\sim8\%$. Therefore, we can conclude that a coarser temporal resolution may lead to lower subhalo lifetimes, but the difference between filament and no filament samples remains within the same significance.
}

{
Another feature of the figure is that differences are noisier for longer lived subhalos. This is the result of snapshot being spaced by scale factor. In this case the error in the accretion time at earlier epochs is larger, i.e. at $a\sim1$ a $\Delta a=0.002$ translates to $\Delta t=29.2Myr$, while at $a\sim0.5$ a $\Delta a=0.002$ translates to $\Delta t=82Myr$.
}

\section{Resolution effects: Subhalo Mass}

\begin{figure}
\begin{center}
\includegraphics[width=.95\linewidth,angle=0]{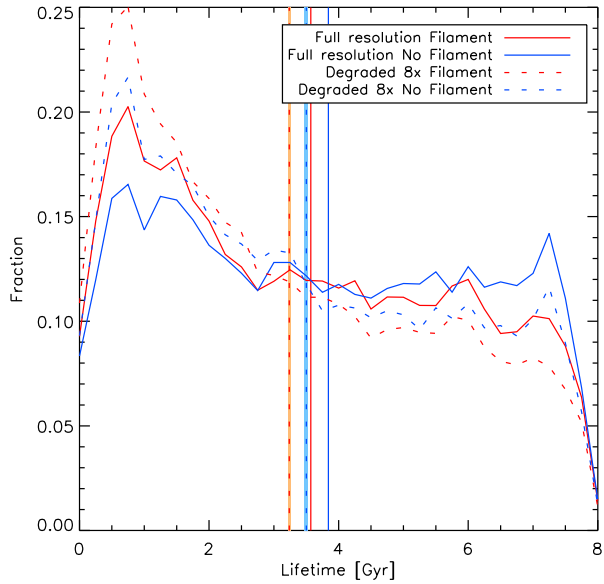}
\caption{
\label{figrefer3}
{ 
Lifetime distribution shown in Figure \ref{fig2b}, along with the results obtained using the $8x$ degraded mass resolution.
Subhalos in the degraded resolution show overall shorter lifetimes. However, the fractional difference between filament and no filament samples is the same as in the full resolution case.
}
}
\end{center}
\end{figure}

{
The mass resolution of the simulation affects the number of resolved subhalos and the mass limit to which we are able to track a subhalo as it loses mass due to dynamical friction. We partially address the mass resolution effect in Figure \ref {fig3} and Table \ref{table2}, where the lifetime differences for more massive subhalos falling from filaments and the field become even larger.  In simulations with coarser resolution our result becomes only stronger, in particular for more massive subhaloes.
}
{
However, to fully address how mass resolution affects lifetimes it is not enough to just use a higher cut on subhalo mass at accretion. Instead, we need to test on a coarser resolution simulation, or include the effect of mass resolution limit in the orbit tracking. 
We chose the latter and, accordingly, for each subhalo orbit we track the subhalo until it reaches $8$ times the original simulation subhalo  mass detection limit, while adding a random scatter corresponding to having $8$ times less particles per subhalo. This model provides a good approximation to the alternative of running a coarser resolution simulation.

Figure \ref{figrefer3} shows the lifetime distribution shown already in Figure \ref{fig2b} but now also including the coarser mass resolution case.
In the coarser case subhalos have lifetimes $3.24\pm0.03$Gyr for the filament sample and $3.50\pm 0.02$Gyr for the no filament samples, both shorter than on the fiducial simulation.  However the fractional difference between these two samples remain at $\sim 8\%$, consistent with the full resolution case.

We conclude that degraded mass resolution leads to lower overall lifetimes, but the fractional difference between filament and no filament samples is not strongly affected leaving our conclusions unchanged.
}

\bibliography{satbib}

\end{document}